\documentclass[12pt]{article}

\def\be{\begin{equation}}
\def\ee{\end{equation}}
\def\bea{\begin{eqnarray}}
\def\eea{\end{eqnarray}}

\def\e{{\rm e}}
\def\tr{{\rm tr}}

\def\d{{\rm d}}
\def\Bbbone{\hbox{{\small $1$} \hspace{-1em} {\normalsize $1$}}}
\def\dis{\displaystyle}

\usepackage{graphicx}
\usepackage{bm}
\usepackage{amsmath}
\usepackage{amssymb}

\begin{document}

\begin{flushright}
hep-th/0611225
\end{flushright}

\pagestyle{plain}

\begin{center}
\vspace{1.5cm} {\Large {\bf Black-Body Radiation Of Noncommutative \\ ~ \\ Gauge Fields}}

\vspace{1cm}

Amir H. Fatollahi$^{~(1}$  \hspace{3mm} and \hspace{3mm} Maryam Hajirahimi$^{~(2}$

\vspace{.5cm}

{\it 1) Mathematical Physics Group, Department of Physics, Alzahra University, \\ P. O. Box 19938, Tehran 91167, Iran}

\vspace{.3cm}

{\it 2) Institute for Advanced Studies in Basic Sciences (IASBS),\\
P. O. Box 45195, Zanjan 1159, Iran}

\vspace{.3cm}

\texttt{fatho@mail.cern.ch\\
rahimi@iasbs.ac.ir}

\vskip .5 cm
\end{center}

\begin{abstract}
The black-body radiation is considered in a theory with noncommutative electromagnetic fields; that
is noncommutativity is introduced in field space, rather than in real space.
A direct implication of the result on Cosmic Microwave Background map is argued.
\end{abstract}

\vspace{1cm}

\noindent {\scriptsize PACS No.: 11.10.Nx, 11.10.Wx, 98.70.Vc}

\newpage

In the recent years much attention has been paid to the formulation
and study of field theories on noncommutative
spaces. The motivation refers to the natural appearance
of noncommutative spaces in some areas of physics, the recent
one in string theory. It has been understood that string
theory is involved by some kinds of noncommutativities;
for examples:
(1) the coordinates of bound states of $N$ D-branes are
represented by $N \times N$ Hermitian matrices \cite{9510135}, and
(2) the longitudinal directions of D-branes in the presence
of a B-field background appear to be noncommutative, as seen by the
ends of open strings \cite{9908142}. In the latter case, we encounter
the spacetime in which the coordinates satisfy the canonical relation
\bea\label{algebra}
[\widehat{x}^{\,\mu},\widehat{x}^{\,\nu}]=i\lambda^{\mu\nu},
\eea
where $\lambda^{\mu\nu}$ is an antisymmetric constant tensor with dimension (length)$^2$.
It is understood that field theories on noncommutative spacetime are defined by actions that
are essentially the same as in the ordinary spacetime, with the exception that the products
between fields are replaced by the $\star$-product, defined for two functions $f$ and $g$ \cite{reviewnc}
\bea
(f\star g)(x)=\exp\big(\frac{i\lambda^{\mu\nu}}{2}\partial_{x_\mu}\partial_{y_\nu}\big)f(x)g(y)\mid_{y=x}
\eea
The pure gauge field sector of noncommutative U(1) theory is defined by the action
\bea
S_{\rm gauge-field}=-\frac{1}{4}\int {\rm d}^{4}x \;\widehat{F}_{\mu\nu}\star \widehat{F}^{\mu\nu}=-\frac{1}{4}\int
{\rm d}^{4}x \;\widehat{F}_{\mu\nu}\widehat{F}^{\mu\nu}
\eea
with $\widehat{F}_{\mu\nu}=\partial_{\mu}A_{\nu}-\partial_{\nu}A_{\mu}-ie[A_{\mu},A_{\nu}]_{\star}$,
by definition $[f,g]_{\star}=f\star g-g\star f$.
The action above is invariant under local gauge symmetry transformation
\bea\label{trans}
A'_{\mu}= U\star A_{\mu}\star U^{-1}+\frac{i}{e}U\star \partial _{\mu}U^{-1}
\eea
in which $U=U(x)$ is the $\star$-phase, defined by a function $\chi(x)$ via the $\star$-exponential:
\bea\label{starphase}
U(x)=\exp_{\star}(i\chi)=1+i\chi-\frac{1}{2}\chi\star\chi+\cdots,
\eea
with $U^{-1}=\exp_{\star}(-i\chi)$, and $U\star U^{-1}=U^{-1}\star U=1$.
Under above transformation, the field strength transforms as
$\widehat{F}_{\mu\nu}\to \widehat{F}^{\prime}_{\mu\nu}=U\star \widehat{F}_{\mu\nu}\star U^{-1}$.
We mention that the transformations of gauge field as well as the field strength look like those of non-Abelian gauge theories.
Besides we mention that the action contains terms which are responsible for interaction between the gauge particles. We see how
the noncommutativity of coordinates induces properties on fields and their transformations, as if they were belonged to a non-Abelian
theory; the subject that how the characters of coordinates and fields may be related to each other is discussed in \cite{fath}.

In a recent work the present authors addressed the radiation we expect from a black-body
in noncommutative space \cite{amma}, in the framework of finite temperature field theory
\cite{fin-temp-book}. As mentioned, noncommutative U(1) gauge theory is involved by
self-interaction of photons, and so beyond the free theory one finds deviations from
the expression by ordinary U(1) theory for black-body radiation. The result by first
loop correction, for $U(T,\Omega)$ as the energy-density receiving from the solid-angle
$\d \Omega$ and for spatial noncommutativity ($\lambda^{0\,i}=0$) is \cite{amma}:
\bea\label{fin-expr}
U(\Omega, T)\,\d \Omega=\Big[\frac{\sigma_0}{4\pi} T^4-\frac{7\pi^4}{675}\,\alpha \,T^4(\lambda T^2)^2 \sin^2\!\theta + {\rm O}\big((\lambda T^2)^4\big)\Big]\d\Omega
\eea
in which $\sigma_0=\frac{\pi^2}{15}$ (Stefan's constant=$\frac{\pi^2}{60}$ for $\hbar=c=k_{\rm B}=1$),
$\alpha=\frac{e^2}{4\pi}\simeq \frac{1}{137}$, $T$ is the temperature, and $\lambda$ is
the length of the vector defined by $\lambda_k=\frac{1}{4}\epsilon_{ijk}\lambda^{ij}$.
The angle $\theta$ is measured from the vector $\bm{\lambda}$ for which we have $0\leq \theta\leq\pi$.

In the present work we address the radiation we expect from a black-body in a theory that
noncommutativity is introduced in field space, rather than space itself. This kind of noncommutativity has
been considered in \cite{nc-field-1}. A U(1) gauge theory with noncommutative fields was introduced
in \cite{nc-field-2}, which is our starting point here. Some other aspects and implications of the model
are studied in \cite{nc-field-3}. Noncommutative gauge fields are introduced by \cite{nc-field-2} ($i,j=1,2,3$)
\bea\label{alfi}
[A_i(x),A_j(y)]=i\,\ell_{ij}\, \delta^{(3)}(\mathbf{x}-\mathbf{y}),\
\eea
together with $[A_i(x),F_j(y)]=i\delta_{ij}\, \delta^{(3)}(\mathbf{x}-\mathbf{y})$, and $[F_i(x),F_j(y)]=0$, in which
$F_i(x)$'s are the conjugate momenta. We mention that, due to presence of $\delta^{(3)}(\mathbf{x}-\mathbf{y})$ in
right-hand side, $\ell_{ij}$'s have the dimension of length. One may add an auxiliary field $A_0$, together with
definitions $F_i=F_{0i}$, $f_{\mu\nu}=\partial_\mu A_\nu - \partial_\nu A_\mu$ ($\mu,\nu=0,1,2,3$),
and $F_{ij}=f_{ij}$. Then by these all, the closest action to the Maxwell's one can be written down as
\cite{nc-field-2}
\bea
S=\int \d^4x \Big( -\frac{1}{4} f^{\,ij}f_{ij}-\frac{1}{2}F^{\,0i}f_{0i}\Big)
\eea
\noindent in which
\bea
F^{\,0i}=\bigg(\frac{\Bbbone}{\Bbbone+ L \, \partial_t}\bigg)^i_{\,\,j}\;f^{\,0j},
\eea
\noindent where $\Bbbone$ is $3\times 3$ identity matrix, and the antisymmetric matrix $L$ is defined by its elements
$L^i_{\,\,j}=\ell_{\,ij}$. We mention that the action is invariant under gauge transformations $A_\mu\to A_\mu-\partial\Lambda$.

By proper choice of coordinates, one always can bring $L$ to the form
\bea
L=\begin{pmatrix}
0 & \ell & 0 \cr -\ell & 0 & 0 \cr 0 & 0 & 0
\end{pmatrix}
\eea
\noindent for which we can show
\bea
\frac{\Bbbone}{\Bbbone+ L \, \partial_t}=\Bbbone -
\frac{\ell\,\partial_t}{1+\ell^2 \partial^2_t}
\begin{pmatrix}
0 & 1 & 0 \cr -1 & 0 & 0 \cr 0 & 0 & 0
\end{pmatrix}
-\frac{\ell^2\,\partial^2_t}{1+\ell^2 \partial^2_t}
\begin{pmatrix}
1 & 0 & 0 \cr 0 & 1 & 0 \cr 0 & 0 & 0
\end{pmatrix}
\eea
As mentioned the action possesses a gauge symmetry, and so we need to add gauge fixing term for
doing calculations, getting
\bea
S+S_{\rm g.f.}=\int \d^4x \Big( -\frac{1}{4} f^{\,ij}f_{ij}-\frac{1}{2}F^{\,0i}f_{0i} -\frac{1}{2} \big(\partial_\mu A^\mu\big)^2\Big)
\eea
By the Fourier expansion of gauge fields
\bea
A_\mu(x)=\frac{1}{(2\pi)^2} \int \d^4 k~ a_\mu(k)~ \e^{ik\cdot x},
\eea
\noindent one can bring things in the momentum space, getting
\bea
S+S_{\rm g.f.}=\int \d^4 k \;\;a_\mu(k) D^{\mu\nu} a_\nu(-k)
\eea
\noindent in which $D^{\mu\nu}$'s are elements of
\bea
D=\begin{pmatrix}
\begin{bmatrix} ~~ & & ~~ \cr  & \widetilde{D} & \cr  &  & \end{bmatrix}  \begin{matrix} 0 \cr 0\cr 0 \end{matrix} \cr
\begin{matrix} ~~~~0 & 0 & 0 & k^2 \end{matrix} \end{pmatrix}
\eea
\noindent where
\bea
\widetilde{D}=\begin{pmatrix}
-k^2+\Delta\ell^2k_0^2(k_1^2+k_2^2) & -\Delta\ell k_0^2(\ell k_0k_1-i k_2) & -\Delta\ell k_0^2(\ell k_0k_2+i k_1)  \cr
 & & \cr
-\Delta\ell k_0^2(\ell k_0k_1+i k_2) & k^2+\Delta\ell^2k_0^4 & i\Delta\ell k_0^3\cr
 & & \cr
-\Delta\ell k_0^2(\ell k_0k_2-i k_1) & -i\Delta\ell k_0^3 & k^2+\Delta\ell^2k_0^4
\end{pmatrix}
\eea
\noindent in which $k^2=k_0^2-\mathbf{k}\cdot\mathbf{k}$, and $\Delta=\displaystyle{\frac{1}{1-\ell^2k_0^2}}$.
We mention that by setting $\ell=0$ we get the propagator for ordinary Maxwell theory in Feynman gauge.

As we are dealing with the black-body radiation a natural framework is finite temperature field theory \cite{fin-temp-book}.
Here, following \cite{kapusta} we choose the real-time formulation; not doing the replacement $t\to i\,t$.
So $k_0=2\pi niT$ in which $T$ is the temperature and $n\in \mathbb{Z}$.
As fields appear quadratically in the action, our theory is a free one, and hence the partition-function ${\cal Z}$, accounting the
contribution of ghosts of gauge fixings, is simply given by:
\bea
{\cal Z} \sim   \Big(\prod_{\mathbf{k},n}k^2\Big)\times \big(\det D\big)^{-\frac{1}{2}}
\eea
\noindent in which the first part is for the contribution by ghosts. As $\det D = k^2\cdot\det\tilde{D}$:
\bea
{\cal Z} &\sim&  \Big(\prod_{\mathbf{k},n}k^2\Big)^{\frac{1}{2}}\times \big(\det \widetilde{D}\big)^{-\frac{1}{2}}\nonumber\\
&\sim& \Big( \prod_{\mathbf{k},n}k^2\Big)^{-1} \times \prod_{\mathbf{k},n}
\Big(1+\ell^2k_0^2\Delta\Big[\frac{k_0^2(k_1^2+k_2^2-k_0^2)}{(k^2)^2}-\frac{k_1^2+k_2^2-2k_0^2}{k^2}\Big]\Big)^{-\frac{1}{2}}
\eea
The total free-energy is simply ${\cal F}=-T \log {\cal Z}$, and by the usual replacements $\log (\det) \to \tr(\log)$ and
$\dis{\sum_{\mathbf{k}}\to \int\!\!\frac{\d^3 k}{(2\pi)^3}}$ we arrive at ($\omega^2=\mathbf{k}\cdot\mathbf{k}$)
\bea\label{fe-1}
\!\!{\cal F}(T)\!\!\!\! &=&\!\!\!\!\! \int\!\!\frac{\d^3 k}{(2\pi)^3}\; T\sum_{n\in\mathbb{Z}} \log (k_0^2-\omega^2)\nonumber\\
& +&\!\!\!\frac{1}{2} \int\!\!\frac{\d^3 k}{(2\pi)^3}\; T\sum_{n\in\mathbb{Z}}
 \log\Big(1+\ell^2k_0^2\Delta\Big[\frac{k_0^2(k_1^2+k_2^2-k_0^2)}{(k^2)^2}-\frac{k_1^2+k_2^2-2k_0^2}{k^2}\Big]\Big)
\eea
\noindent For doing the sum on $n$ it is very convenient to use the contour integral formula in complex $k_0$-plane,
that gives for a function $f(k_0)$ \cite{kapusta}
\bea\label{ci}
T\sum_{n\in\mathbb{Z}} f(k_0=2\pi n i T) &=& \frac{1}{2\pi i} \int_{-i\infty + \varepsilon}^{\;i\infty+\varepsilon}
\;\frac{\d k_0}{\exp(\beta k_0)-1} \big[f(k_0) + f(-k_0)\big]\nonumber\\
&& + \frac{1}{2\pi i} \int_{-i\infty}^{\;i\infty} \d k_0 \; f(k_0)
\eea
\noindent in which $\varepsilon>0$ is an infinitesimal, and $\beta=T^{-1}$. The contours in integrals above
usually are closed in $k_0>0$ region and so counterclockwise \cite{kapusta}. We also recall that any quantity in finite temperature
theory is defined after subtraction by its $T=0$ counterpart, and so $T$-independent parts of expressions, infinite or finite, are
not important \cite{fin-temp-book}. For the first term in (\ref{fe-1}) one simply finds:
\bea
\int\!\!\frac{\d^3 k}{(2\pi)^3}\; T\sum_{n\in\mathbb{Z}} \log (k_0^2-\omega^2) \to -\,\frac{\pi^2}{45}\,T^4 + ``\,\infty{\rm ~independent~of~}T\,"
\eea
\noindent which is the standard expression for the free-energy of a black-body \cite{fin-temp-book}. For the remaining part of
(\ref{fe-1}), assuming that the deviation from the standard expression is very slightly, that is assuming $\ell^{-1}$ is practically infinite
for being taken by $k_0$, and so $\ell k_0\ll 1$, we can perform an $\ell$-expansion.  So, by $\log(1+\eta)\sim\eta$ for $\eta\ll 1$, we reach to
\bea
{\cal F}(T) &=& -\frac{\pi^2}{45}\,T^4\! +\frac{\ell^2}{2} \int\!\!\frac{\d^3 k}{(2\pi)^3}\; T\sum_{n\in\mathbb{Z}}
k_0^2\Big[\frac{k_0^2(k_1^2+k_2^2-k_0^2)}{(k^2)^2}-\frac{k_1^2+k_2^2-2k_0^2}{k^2}\Big]\nonumber\\
&& + ``\,\infty{\rm ~independent~of~}T\,"+{\rm O} (\ell^4).
\eea
\noindent Keeping the contribution of the pole $k_0=\omega$ in the contour integral (\ref{ci}), one can do summation on $n$.
With $\mathbf{k}=(\omega, \theta,\phi)$ in spherical coordinates, together with
$\d^3 k = \omega^2\d\omega\d(\cos\theta)\d\phi=\omega^2\d\omega\d\Omega$, we find:
\bea
{\cal F}(T)= -\frac{\pi^2}{45}\,T^4 + \frac{\ell^2}{4} \int\!\!\frac{\d\Omega}{(2\pi)^3}
\cos^2\theta\int_0^\infty\d\omega \frac{\omega^5}{\e^{\beta \omega}-1}\Big(1-\frac{\beta\omega\e^{\beta\omega}}{\e^{\beta \omega}-1}\Big)
+{\rm O} (\ell^4)
\eea
\noindent By using:
\bea
\int_0^\infty\frac{s^{2m +1}\d s}{(\e^{s}-1)}=\zeta(2m +2)(2m +1)!,
\\
\int_0^\infty\frac{s^{2m +2}\e^{s}\d s}{(\e^{s}-1)^2}=\zeta(2m +2)(2m +2)!,
\eea
with $\zeta(t)$ as the Riemann zeta-function, we find:
\bea
{\cal F}(T)=-\frac{\pi^2}{45}\,T^4 - \frac{5\pi^3}{252} \ell^2\, T^6  \int\!\!\d\Omega \cos^2\theta+ {\rm O} (\ell^4)
\eea
As the total energy is given by $U={\cal F}- T\partial_{\,T}{\cal F}$, for
the energy receiving from the solid-angle $\d\Omega$ we find:
\bea\label{fin-expr2}
U(T,\Omega)\,\d \Omega=\bigg[\frac{\sigma_0}{4\pi} T^4
+ \frac{25\pi^3}{252}\, \ell^2 \, T^6 \cos^2\theta \bigg] \,\d \Omega +{\rm O} (\ell^4T^4)
\eea
in which $\sigma_0=\frac{\pi^2}{15}$ (Stefan's constant=$\frac{\pi^2}{60}$).

Although it is hard to imagine that the implications of noncommutativity can be detected in
a laboratory black-body radiation, one may look for an indication of noncommutativity
in the signals we are getting from the extremely hot seconds of early universe.
In fact, the energy scale that one expects for relevance of noncommutative effects is as much as
high that suggests maybe it has been available for particles only in the early universe.
So an excellent way to test the phenomenon related to noncommutativity would be the study of
what are left for us as early universe's heir, the most important among them the Cosmic Microwave Background
(CMB) radiation. The reason is, CMB map is just a tableau of events which happened at the first seconds of
universe, at the decoupling era or much earlier, when the energies were sufficiently high to make relevant
possible spacetime noncommutativity. There have been efforts to formulate and study the
noncommutative versions of inflation theory \cite{nc-cosmos}. In \cite{fuzzy-sphere} by taking the blowing sphere
that eventually plays the role of the so-called last-scattering surface as a fuzzy sphere
some kinds of explanation is presented for the relatively low angular power spectrum $C_l$ in small $l$ region ($l\simeq 6)$.
In \cite{tx-uncer} the consequences of space-time uncertainty relations of the form
$\Delta t \Delta x\geq l_s^2$ are studied in the context of inflation theory, and possible applications of these relations
in better understanding of present CMB data are discussed. As uncertainty relations usually point to noncommutative objects,
these kinds of efforts also might be regarded as studies that are inspecting, though indirectly, the implications of noncommutativity on CMP map.

As CMB map is in fact a black-body radiation pattern which is slightly perturbed by
fluctuations, instead of dealing with the implications of noncommutativity on
different cosmological models, we can directly address what one should expect to see
in CMB map if in early universe the coordinates or fields have satisfied the algebra (\ref{algebra}) or (\ref{alfi}), respectively.
According to the expressions (\ref{fin-expr}) and (\ref{fin-expr2}), space and field noncommutativity in early universe modify
the pattern we expect to see in the CMB map sky. Also, in \cite{nc-field-2} the implication of noncommutative gauge fields on the pattern we expect
to see in polarized CMB data is considered.

Much effort is currently being devoted to examining the CMB temperature anisotropies measured with the Wilkinson Microwave
Anisotropy Probe (WMAP) \cite{wmap1,wmap3}. It would be extremely important if the present and forthcoming data indicate any
significant evidence for canonical noncommutativity in the early universe, the thing that one of its direct implications
is given by an anisotropic radiation.


\end{document}